\documentclass[preprint,review,12pt]{elsarticle}
\usepackage[utf8]{inputenc}
\usepackage{graphicx}
\usepackage{amssymb}

\newcounter{bla}

\journal{Computer Physics Communications}

\begin{document}

\begin{frontmatter}

\title{New magnetohydrodynamic model available at NASA Community Coordinated Modeling Center}
\author[a,b,c]{Ilja Honkonen\corref{author}}
\author[d]{Lutz Rast\"atter}
\author[d]{Alex Glocer}
\cortext[author]{Corresponding author.\\\textit{E-mail address:} ilja.honkonen@fmi.fi}
\address[a]{Department of Physics, Catholic University of America, Washington, DC, USA}
\address[b]{stationed at NASA Goddard Space Flight Center, Greenbelt, MD, USA}
\address[c]{now at Finnish Meteorological Institute, Helsinki, Finland}
\address[d]{NASA Goddard Space Flight Center, Greenbelt, MD, USA}

\begin{abstract}

The Community Coordinated Modeling Center (CCMC) at NASA Goddard Space Flight Center is a multi-agency partnership to enable, support and perform research and development for next-generation space science and space weather models. CCMC currently hosts nearly 100 numerical models and a cornerstone of this activity is the Runs on Request (RoR) system which allows anyone to request a model run and analyze/visualize the results via a web browser. CCMC is also active in the education community by organizing student research contests, heliophysics summer schools, and space weather forecaster training for students, government and industry representatives.

We present a generic magnetohydrodynamic (MHD) model - PAMHD - that has been added to the CCMC RoR system which allows the study of a variety of fluid and plasma phenomena in one, two and three dimensions using a dynamic point-and-click web interface.
Flexible initial and boundary conditions allow experimentation with a variety of plasma physics problems such as shocks, instabilities, planetary magnetospheres and astrophysical systems.
Experimentation with numerical effects, e.g. resolution, solution method and boundary conditions, is also possible and can provide valuable context for space weather forecasters when interpreting observations or modeling results.

We present an overview of the C++ implementation and show example results obtained through the CCMC RoR system, including the first to our knowledge MHD simulation of the interaction of the magnetospheres of Jupiter and Saturn in two dimensions.

\end{abstract}


\end{frontmatter}

{\bf PROGRAM SUMMARY}

\begin{small}
\noindent
{\em Program Title:} PAMHD \\
{\em Licensing provisions:} main program GPLv3, supporting code 3-clause BSD \\
{\em Programming language:} C++11 \\
{\em Nature of problem:} fully ionized plasma modeled as a single magnetohydrodynamic fluid with planetary, magnetospheric and heliospheric applications \\
{\em Solution method:} finite volume method with Godunov type solvers [1], parallelized using the message passing interface [2] and with elliptic cleaning of divergence of magnetic field [3] \\
{\em Additional comments including Restrictions and Unusual features:} simulations can be prepared, executed and visualized through NASA Community Coordinated Modeling Center website, analytic B0+B1 split with curl B0 = 0, Cartesian grid \\
\\

\end{small}

\section{Introduction}
\label{sec:intro}

The Community Coordinated Modeling Center (CCMC) hosts over 100 models in heliospheric, magnetospheric, ionospheric and thermospheric physics that have been developed to solve a specific problem.
While a few models, such as the Space Weather Modeling Framework \cite{toth05}, have over the years been extended from their original domain, no other model at CCMC currently allows as flexible experimentation with plasma physics as PAMHD.
Furthermore, using a publicly accessible repository for simulation results also supports various open access policies implemented on national and international level e.g. in US and EU.
The Community Coordinated Modeling Center is one of the data repositories suggested by the American Geophysical Union for archiving simulation results (http://sites.agu.org/publications/files/2014/06/Data-Repositories.pdf retrieved on 2016-10-09).
Quick look plots enable permanent URL links to simulation results which can be used for many purposes, such as to support teaching a course or to create a collection of results on a particular phenomenon for easy access.

We present a freely available generic program for simulating plasma using the magnetohydrodynamic (MHD) approach that anyone can download, use, study, modify and redistribute.
The code is available at https://github.com/iljah/pamhd/tree/mhd and
the majority of its functionality is available through the NASA CCMC website at http://ccmc.gsfc.nasa.gov/models/modelinfo.php?model=PAMHD where users can view results from existing runs and submit new ones using any web browser with JavaScript support.

In Section \ref{sec:features} we present the most important features of the model and in Section \ref{sec:overview} we show a high-level overview of its implementation.
We present a variety of results obtained with the model through CCMC in Section \ref{sec:results} and draw out conclusions in Section \ref{sec:conclusions}.

\section{Model features}
\label{sec:features}

Here we present the MHD part of PAMHD which solves equations of ideal magnetohydrodynamics in conservative form in one to three dimensions.
Currently, a finite volume method is used with Godunov type first order solvers \cite[e.g.][]{leveque02}.
Divergence of cell-centered magnetic field is removed with the elliptic cleaning method of \cite{brackbill80}.
Parallel computation with MPI is achieved by using the DCCRG library \cite{honkonen13} but support for adaptive mesh refinement provided by DCCRG is not used here.
The JavaScript Object Notation (JSON) format \cite{rfc7159} is used for configuration files which allows them to be easily manipulated both manually and using almost any programming language.
Normalization of physical quantities can also be customized by specifying the adiabatic index, vacuum permeability and particle mass used by the model.

\begin{figure}
\includegraphics[width = 0.75 \columnwidth]{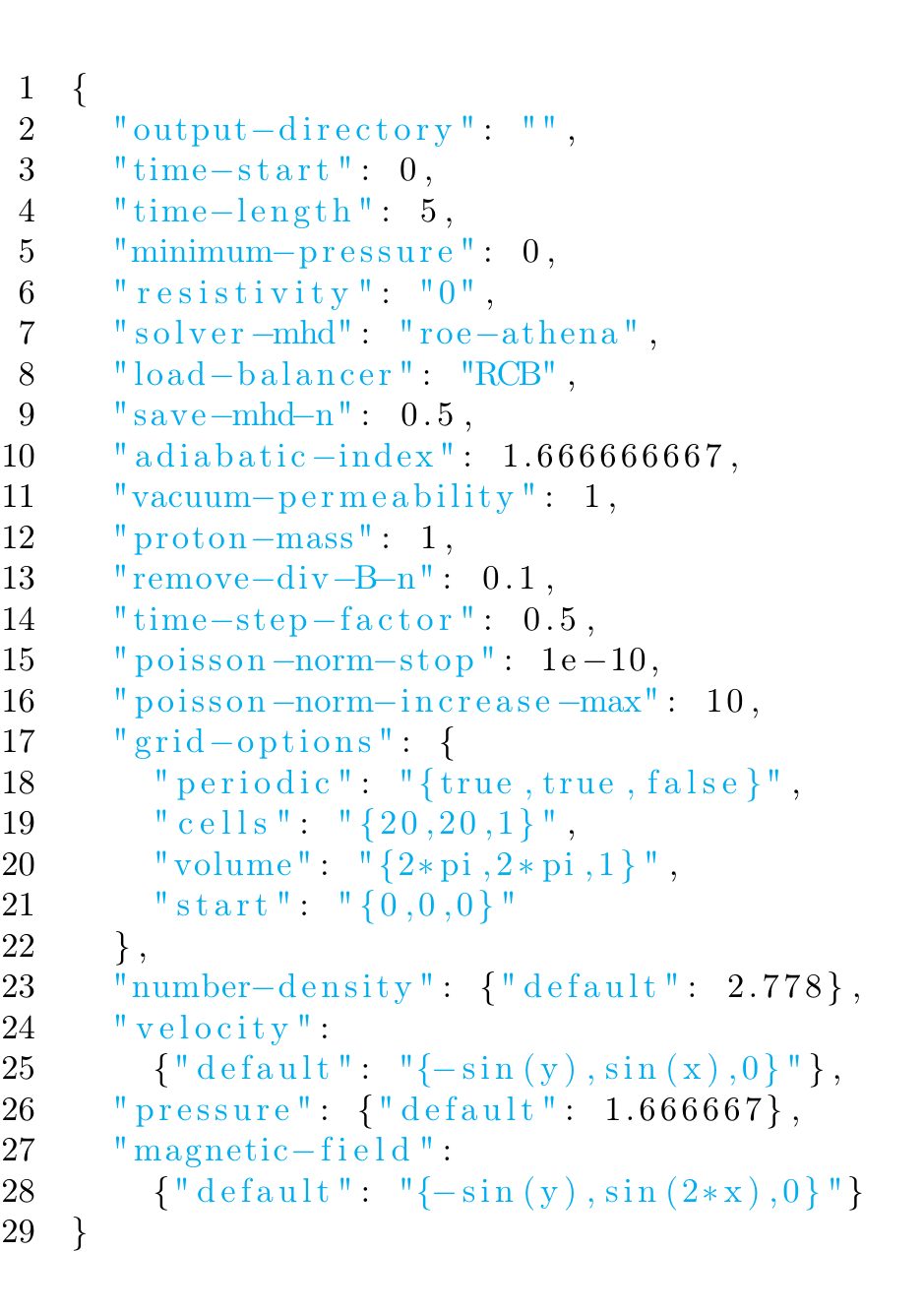}
\caption{
Configuration file for an Orszag-Tang vortex simulation in JSON format.
}
\label{fig:orszagtangcfg}
\end{figure}

\begin{figure}
\includegraphics[width = 0.75 \columnwidth]{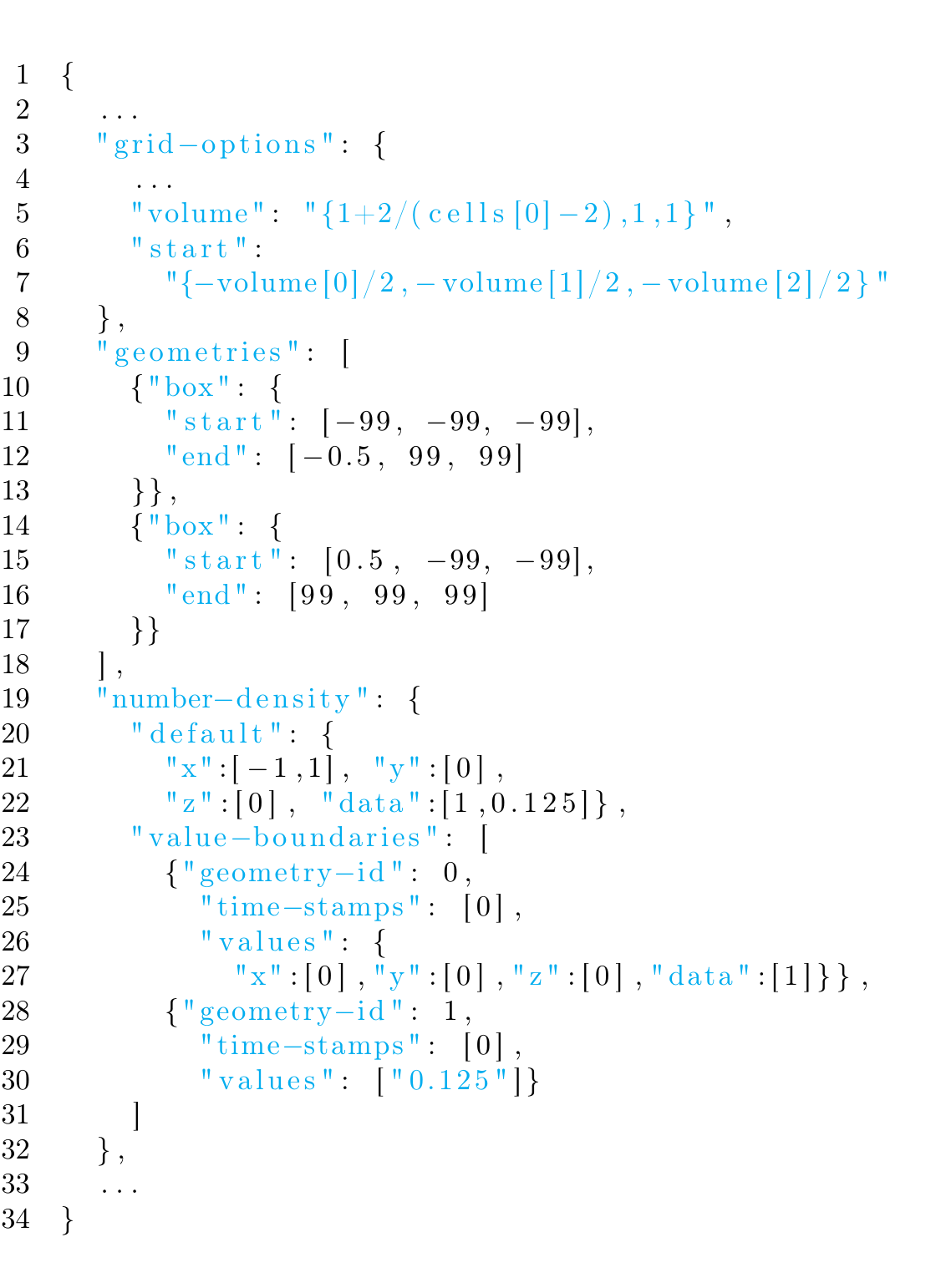}
\caption{
Partial configuration file for a shock tube simulation.
Lines 2 and 4 represent the same configuration as in Figure \ref{fig:orszagtangcfg}, while line 33 represents the configuration for plasma velocity, pressure and magnetic field.
}
\label{fig:shocktubecfg}
\end{figure}

Figure \ref{fig:orszagtangcfg} shows a full configuration file for the Orszag-Tang vortex test.
Basic simulation parameters, physical constants, etc. are given on lines 2..16.
Lines 17..22 specify the simulation grid, i.e. the number of cells in each dimension, whether the system is periodic in any dimension and the simulation volume and its starting coordinate.
Grid parameters are mathematical expressions and the expression for volume can refer to the cells variable and the start expression can additionally refer to volume variable as shown in Figure \ref{fig:shocktubecfg}.
In the simplest case, the configuration for plasma parameters consists of their default values with which the simulation is initialized.
A value can be either:
1) a number, or a JSON array of numbers in case of vector variables, to which the simulation variable is set in all simulation cells, e.g. number density on line 23 in Figure \ref{fig:orszagtangcfg}.
2) a string representing a mathematical expression, enclosed in \{\} in case of vector variables, that can refer to the center of current cell with either $x, y, z$ or $radius, lat, lon$ which are evaluated for each cell, e.g. velocity on lines 24..25.
3) a homogeneous Cartesian or spherical grid of values given as a JSON object with keys ``data" and ``x", ``y", ``z" or ``radius", ``lat", ``lon" whose corresponding values are arrays of numbers shown lines 20..22 of Figure \ref{fig:shocktubecfg}.
The number array for ``values" uses x or radius-first ordering i.e. 2nd number corresponds to 2nd coordinate in ``x" and 1st coordinates in ``y" and ``z".
Nearest neighbor interpolation is used in time and space when setting plasma parameters in simulation cells.

Generic initial and boundary conditions allow a wide variety of systems to be modeled.
These are defined separately for each variable, in this case number density, velocity, pressure and magnetic field.
Mathematical expressions can be used to provide space and time-dependence and they can refer to predetermined variables, e.g. x, y and z for current cell's center coordinates, whose value is set at the time of query.
Two types of boundaries are supported: copy and value.
A value boundary sets the corresponding variable in all grid cells within the boundary to given value between every solution.
The value(s) to set in value boundaries are given similarly to default initial condition with the addition that more than one value can be given representing different points in time.
A copy boundary sets the corresponding variable to the average value in all non-boundary face neighbor cells and has no effect if a cell within the boundary only has other boundary cells as face neighbors.
In the special case that all simulation variables of a cell belong to a copy boundary and the cell only has face neighbors of boundary type, the cell is excluded from the solution i.e. it is excluded from removal of divergence of magnetic field and MHD fluxes into and out of that cell are not calculated.
Two geometry types are supported for specifying the volume occupied by a boundary or initial condition: box and sphere.

Figure \ref{fig:shocktubecfg} shows a configuration file of a shock tube test which uses value boundaries with a box geometry to keep the plasma parameters constant at ends of the tube.
Lines 2 and 4 are similar to lines 2..16 and 18..19 respectively in Figure \ref{fig:orszagtangcfg} while line 30 marks the omitted configuration for rest of simulation variables that are similar to lines 19..32.
Line 5 shows an example of the expression for volume referring to the cells variable while line 7 shows expression for start referring to volume.
Two box geometries are specified at ends of the tube on lines 10..17 for keeping plasma parameters constant during the simulation.
The value boundaries on lines 22..27 are set to locations of above geometries using "geometry-id" as the key and the index of the geometry in the geometry list as the value.
As both value boundaries are constant in time only one time stamp and value is used.

\section{Code overview}
\label{sec:overview}

PAMHD is written in C++11, the version of C++ language standardized in 2011, because of added support for passing an arbitrary number of template arguments to classes and functions - variadic templates.
This allows great flexibility in e.g. code coupling as shown in \cite{honkonen15} with the generic simulation cell class that provides tagged simulation variables similar to tagged dispatch of functions.
Tagged variables are referred to, i.e. their data is accessed by, a tag which allows new variables to be added or old ones removed without modification of existing code as opposed to using hard-coded identifiers as done e.g. in \cite{einkemmer16}.
At best several distinct computational models can be combined without modifying existing code to run simultaneously in the same volume \cite{honkonen15}.
The generic simulation cell class also provides support for Message Passing Interface (MPI) standard by allowing the transfer of one or more simulation variables between processes to be switched on or off either globally or on a cell-by-cell basis.
This feature plays well with the DCCRG library used by PAMHD \cite{honkonen13}, which encapsulates the details of nearest-neighbor communication, adaptive mesh refinement, domain decomposition, etc.
The HLL, HLLD and Roe MHD solvers used by PAMHD have been borrowed from Athena \cite{stone08}.

\begin{figure}
\includegraphics[width = 0.75 \columnwidth]{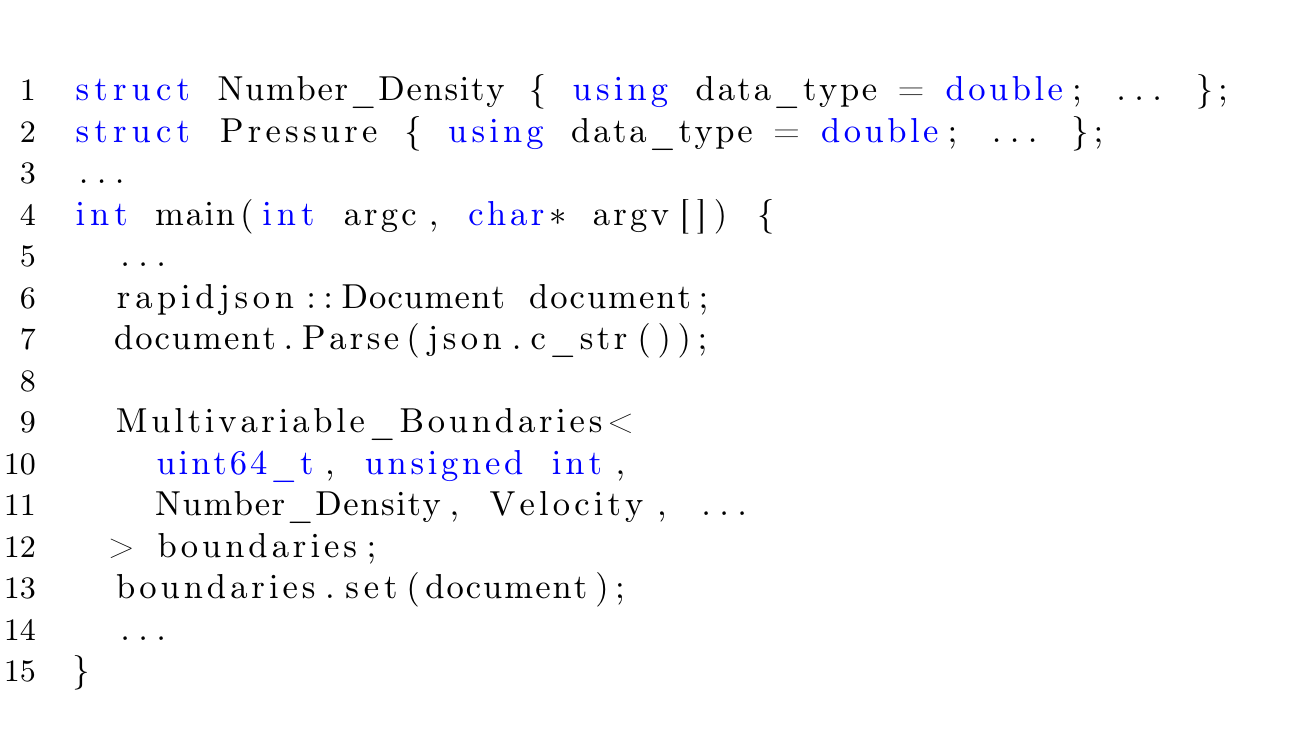}
\caption{
Relevant details of the support for an arbitrary number of initial and boundary condition variables implemented using variadic templates.
Lines 1..3 show the variables, which are given to the boundary super-class on line 11.
The configuration file is parsed on lines 6..7.
On line 13 the boundary super-class goes through all variables (given to it on line 11) and prepares the corresponding initial and boundary conditions from the given configuration.
}
\label{fig:bdyex}
\end{figure}

The generic initial and boundary conditions discussed in Section \ref{sec:features} also make use of tagged simulation variables as illustrated in Figure \ref{fig:bdyex}.
Before the simulation is started, and after every rebalance of computational load i.e. domain decomposition if using adaptive mesh refinement, simulation cells must be classified into normal, boundary and do-not-solve cells based on boundaries specified in the configuration file.
In order to allow greater flexibility in specifying boundary conditions, a particular boundary does not have to modify every plasma parameter within its volume i.e. one simulation cell can be a boundary condition for one plasma parameter while being a normal cell for another parameter.
Since boundary classification and other logic must be applied separately for every boundary variable it has been generalized via variadic templates.
The boundary super-class used by the main program is given, as template parameters, all boundary variables of interest.
When called from the main program, functions of the boundary super-class are applied recursively at compile-time to every boundary variable.
In the example of Figure \ref{fig:bdyex} boundary variables are defined on lines 1..2 and the boundary condition on lines 9..12 with an arbitrary list of boundary variables given on line 11.
The configuration file is parsed on lines 6..7 and given to the boundary class on line 13 which applies the set function, used to create a simulation boundary for one variable, to every boundary variable from line 11.

\subsection{Dynamic run submission page}

The ability of PAMHD to have an arbitrary number of initial and boundary conditions makes it difficult to use a static web page for submitting run requests, which has mostly been the case so far at CCMC.
Therefore a new run submission page was developed that allows geometries, initial and boundary conditions to be added and removed on client side using JavaScript (JS) without multiple HTTP requests between client and server.
Currently the run submission page does not use external JS libraries.
The configuration file creator can be used locally by downloading the gh-pages branch of PAMHD or online at https://nasailja.github.io/pamhd/configuration\_creator/mhd/index.html.
The CCMC run submission page uses a customized version of above page.

The top of the run submission page contains buttons for example simulations which, when clicked, fill out the page automatically.
These presets should allow users to get started with the configuration creator easily via several examples with varying complexity.
It is also possible to fill out the configuration from an existing run by clicking the "Import configuration from existing run" button, which will download a list of existing runs, and selecting the desired run id.
Due to security specification of HyperText Markup Language (HTML) the import button only works from the CCMC version of the run submission page.
The resulting JSON configuration file can also be viewed by clicking the "View configuration" button, and copied and pasted into a file for running a locally installed version of the model.
The preset buttons at top of the page also update the configuration file view.

The configuration is created from mandatory simulation parameters, physical constants, etc. as well as an arbitrary number of initial and boundary conditions.
Each initial and boundary condition is associated with a geometry and is applied to cells within that geometry.
The initial and boundary conditions of each plasma parameter (number density, velocity, pressure and magnetic field) as separate from each other.
Rusanov or HLL solvers also support multiple curl-free background magnetic dipole fields that can be entered in the configuration interface.
Geometries can be added by clicking the "Add box" and "Add sphere" buttons while the last added geometry can be removed by clicking "Remove last" button.
Initial conditions, value and copy boundaries can similarly be added to each variable and removed by clicking the respective buttons.
In this case the geometry to which the boundary in question applies must be specified, the ids are listed next to each geometry.

\begin{figure}
\includegraphics[width = 0.75 \columnwidth]{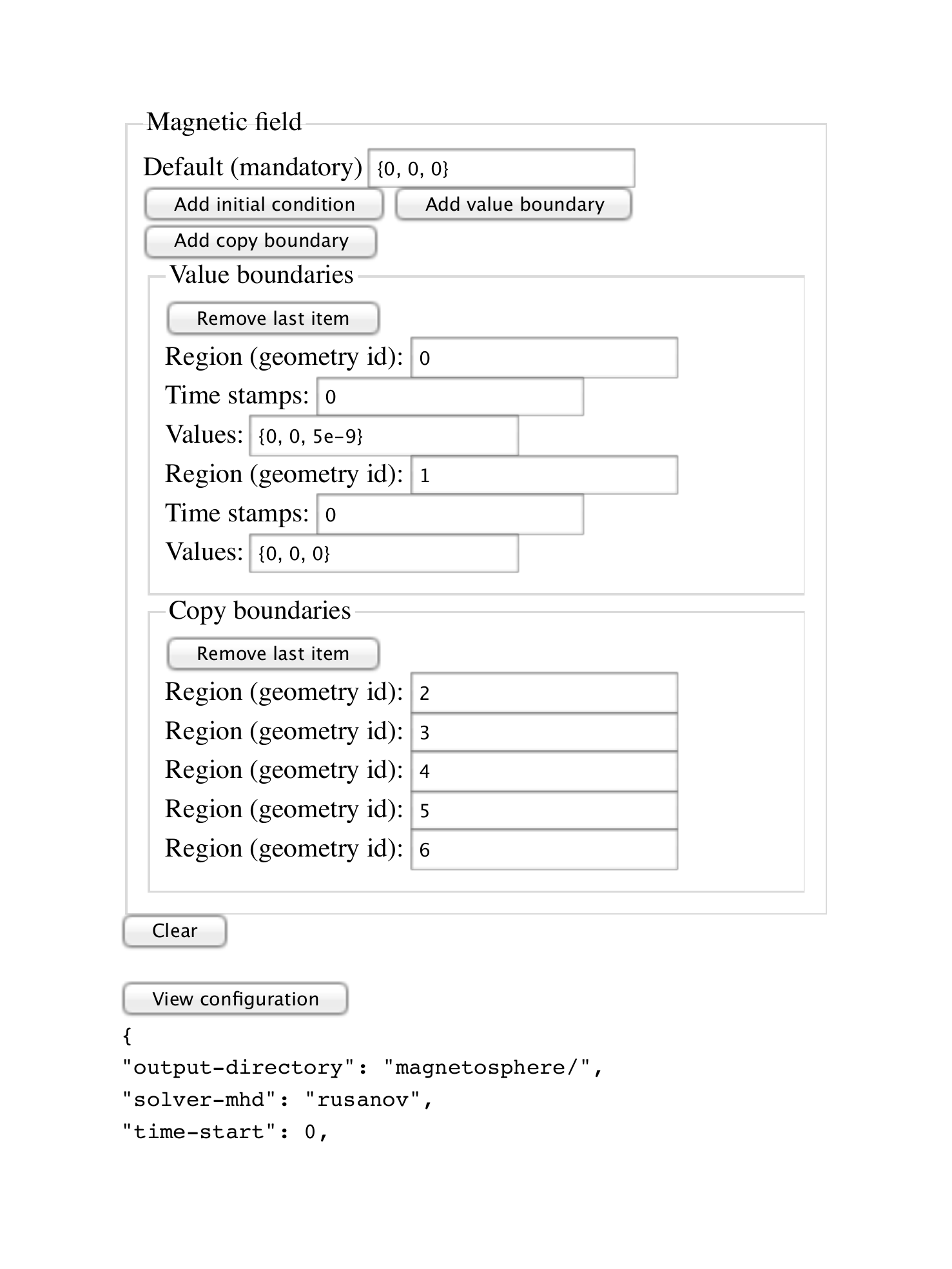}
\caption{
Excerpt from run configuration page after clicking "Earth's magnetosphere" preset button.
}
\label{fig:html}
\end{figure}

Figure \ref{fig:html} shows an excerpt from the web page
after clicking the "Earth's magnetosphere" preset button.
The run setup consists of 5 copy boundaries at faces of simulation box perpendicular to Sun-Earth line and the face at the anti-Sunward side of the box.
The value boundary representing solar wind input is located in geometry id 0 and the value boundary representing inner boundary of the magnetosphere is located in geometry id 1.
The beginning of this configuration's configuration file is also visible at the bottom of the figure.

\section{Example run requests and results}
\label{sec:results}

The model description page at http://ccmc.gsfc.nasa.gov/models/modelinfo.php?model=PAMHD provides links to the database of existing runs and the run submission page where new runs can be requested.
Links to instructional videos of requesting new runs and visualizing results are also provided.
Here we show example results in 1, 2 and 3 dimensions starting from local or MHD-in-a-box simulations and conclude with the first to our knowledge MHD simulation of the interaction of magnetospheres of Jupiter and Saturn in two dimensions.

\begin{figure}
\includegraphics[width = \columnwidth]{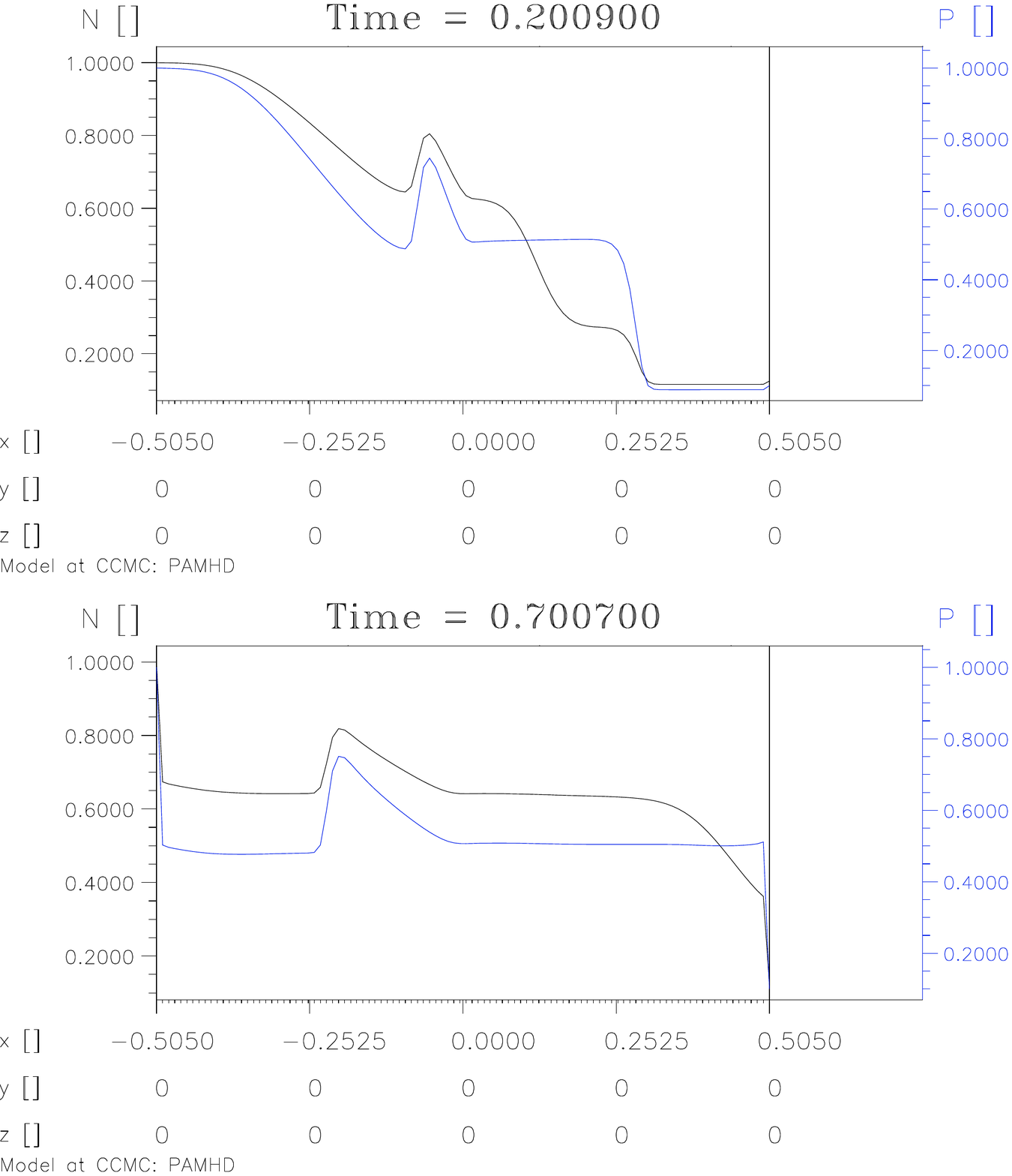}
\caption{
Shock tube simulation plotted from http://ccmc.gsfc.nasa.gov/results/viewrun.php?domain=LP\&runnumber=ilja\_honkonen\_20161017\_LP\_1
}
\label{fig:shocktube}
\end{figure}

Figure \ref{fig:shocktube} shows the solution to a shock tube type problem which is used extensively for benchmarking solvers \cite[e.g.][]{sod78}.
We use Roe's solver \cite{roe81} with $100+2$ cells and show the result at time 0.2 and the interaction of waves from the discontinuity with value boundaries at ends of the tube at time 0.7.
Initial left and right states are $\rho = 1, P = 1, B_y = 1$ and $\rho = 0.125, P = 0.1, B_y = -1$ respectively with $\vec{V} = \vec{0}, B_x = 0.75, B_z = 0$ everywhere and $\mu_0 = 1, \gamma = 5/3, m_p = 1, CFL = 0.5$.
This example reproduces the "Shock tube" preset from the run submission page and is available as run ilja\_honkonen\_20161017\_LP\_1 on the CCMC database query website http://ccmc.gsfc.nasa.gov/ungrouped/LP/LP\_db.php.

\begin{figure}
\includegraphics[width = 0.75 \columnwidth]{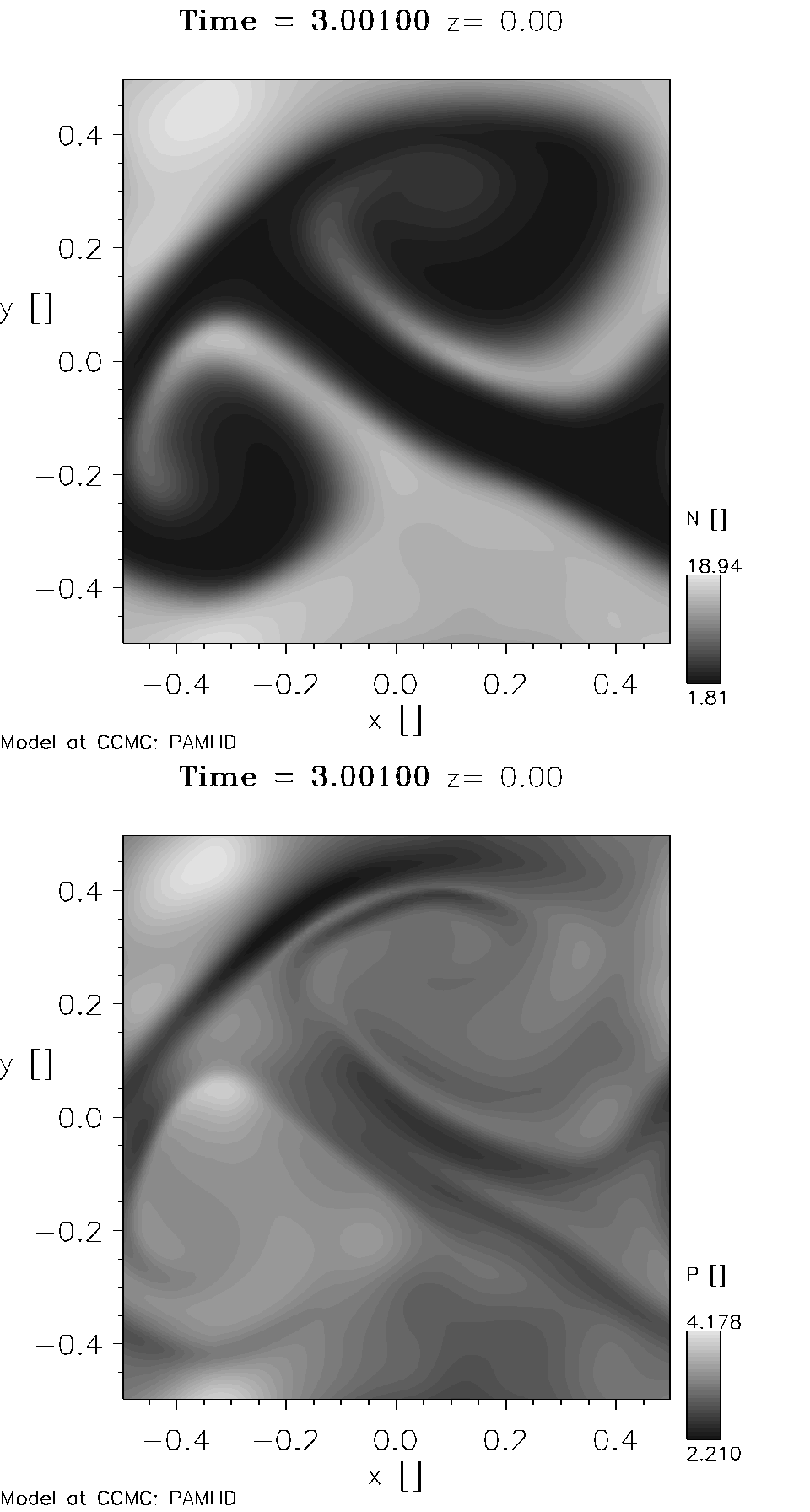}
\caption{
Kelvin-Helmholtz instability simulation plotted from http://ccmc.gsfc.nasa.gov/results/viewrun.php?domain=LP\&runnumber=ilja\_honkonen\_20161017\_LP\_2
}
\label{fig:KH}
\end{figure}

The Kelvin-Helmholtz instability is often used for testing numerical solvers but it is also an important driver of space weather, for example, via plasma entry from solar wind into magnetosphere and driving of ultra-low-frequency waves which in turn strongly affect the radiation belts \cite{wu86, kavosi15}.
Figure \ref{fig:KH} shows density and pressure in a Kelvin-Helmholtz simulation at 3 s using Roe's solver with 50x50 cells in a volume $|x| <= 0.5, |y| <= 0.5$, periodic grid and $\mu_0 = 1, \gamma = 1.4, m_p = 1, CFL = 0.5$.
Initial condition is $|y| < 0.25: \rho = 2, V_x = -0.5$ and $|y| >= 0.25: \rho = 15, V_x = +0.5$ and additionally $V_x = 0.1~sin(2 \pi x), V_y = 0.1~cos(2 \pi x)$ and $P = 3, B_x = 0.25$ everywhere.
Diverge of magnetic field is cleaned every 0.05 seconds.
This example reproduces the "Kelvin-Helmholtz" preset from the run submission page with higher resolution and is available as run ilja\_honkonen\_20161017\_LP\_2.

\begin{figure}
\includegraphics[width = \columnwidth]{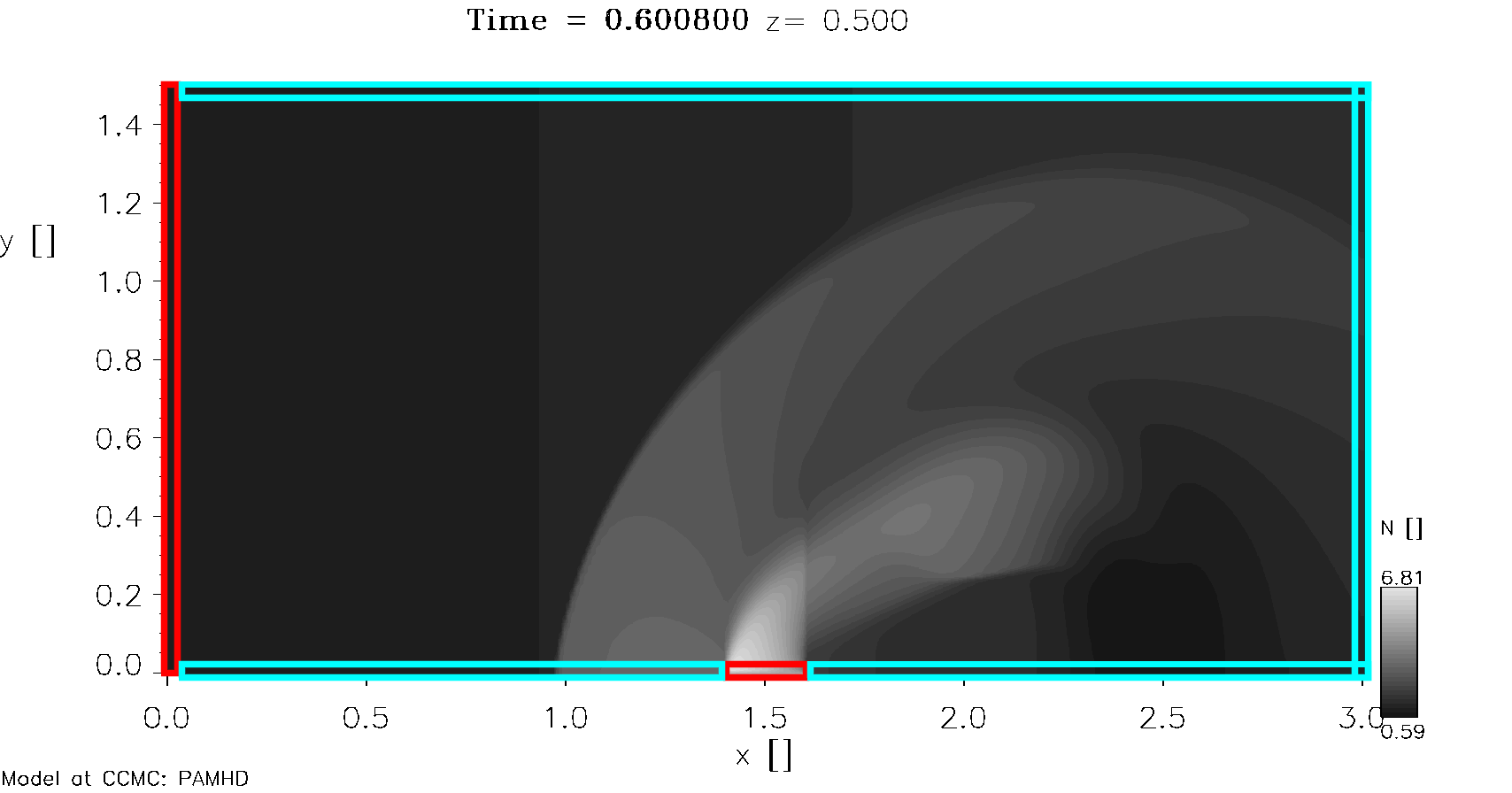}
\caption{
Density in a jet injection simulation with positions of value and copy boundaries indicated in red and light blue respectively.
}
\label{fig:annotated}
\end{figure}

\begin{figure}
\includegraphics[width = \columnwidth]{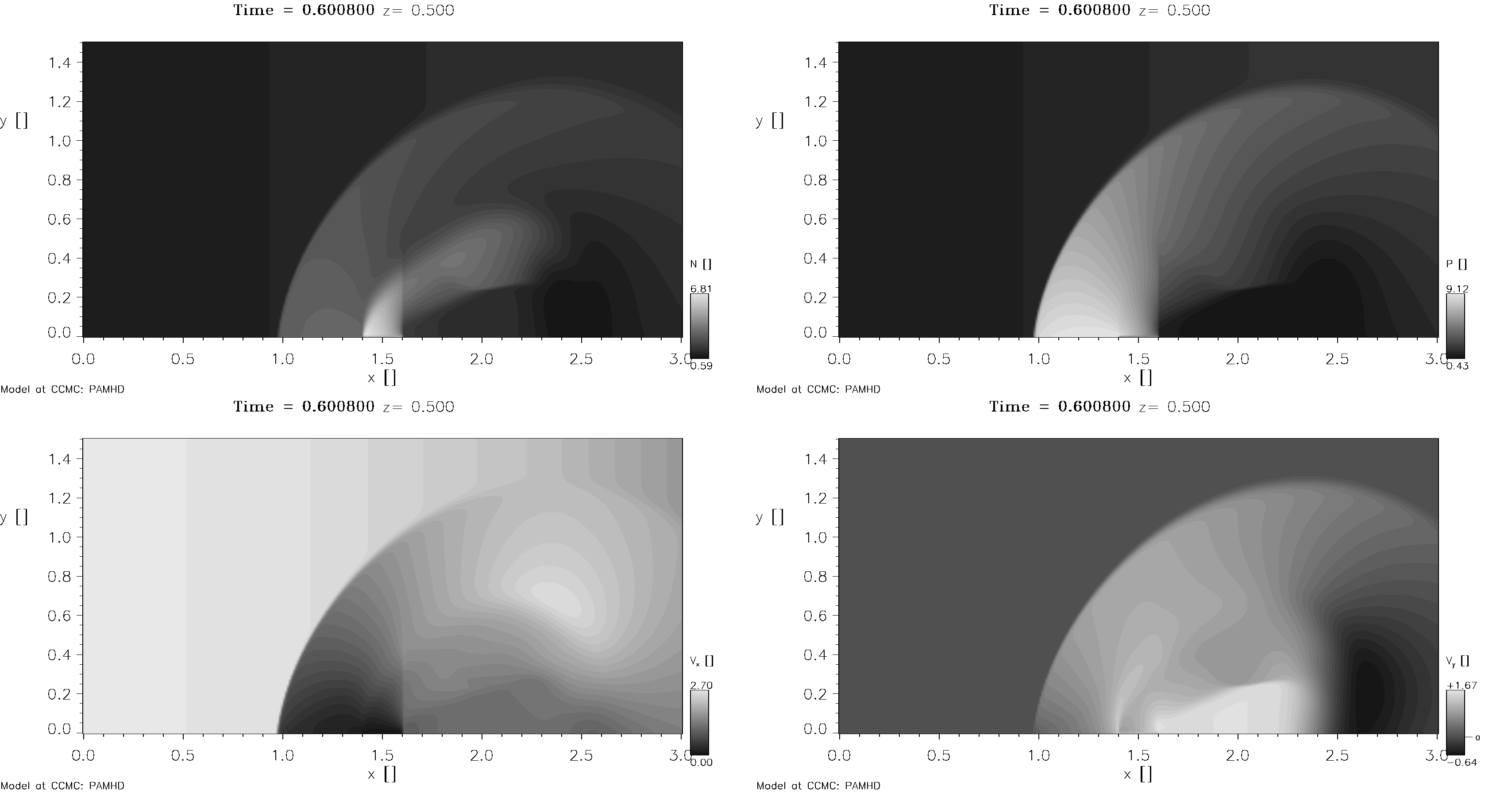}
\caption{
Density, pressure and in-plane components of velocity in jet injection simulation plotted from http://ccmc.gsfc.nasa.gov/results/viewrun.php?domain=LP\&runnumber=ilja\_honkonen\_20161017\_LP\_3
}
\label{fig:jet}
\end{figure}

Figure \ref{fig:annotated} shows the density in a jet injection simulation similar to \cite{suarez12} at time 0.6, illustrating interleaved value and copy boundaries along edges of the simulation box highlighted in red and light blue respectively.
The result is obtained using Roe's solver with 40x20 cells over a volume 3x1.5 not including one layer of boundary cells.
Initial condition is $\rho = 1, \vec{V} = (2.7-x/2,0,0), P = 1$ with inflow from left boundary of $\rho = 1, \vec{V} = (2.7,0,0), P = 1$ and inflow from bottom at $1.4 < x < 1.6$ of $\rho = 5, \vec{V} = (0,1.3,0), P = 5$, $\mu_0 = 1, \gamma = 5/3, m_p = 1, CFL = 0.5$ and $\vec{B} = \vec{0}$ everywhere.
Figure \ref{fig:jet} shows density as well as pressure and in-plane components of velocity with 32 contour level in a 300x150 cell simulation excluding a layer of boundary cells.
This example reproduces the "Jet injection" preset from the run submission page with higher resolution and is available as run ilja\_honkonen\_20161017\_LP\_3.

\begin{figure}
\includegraphics[width = \columnwidth]{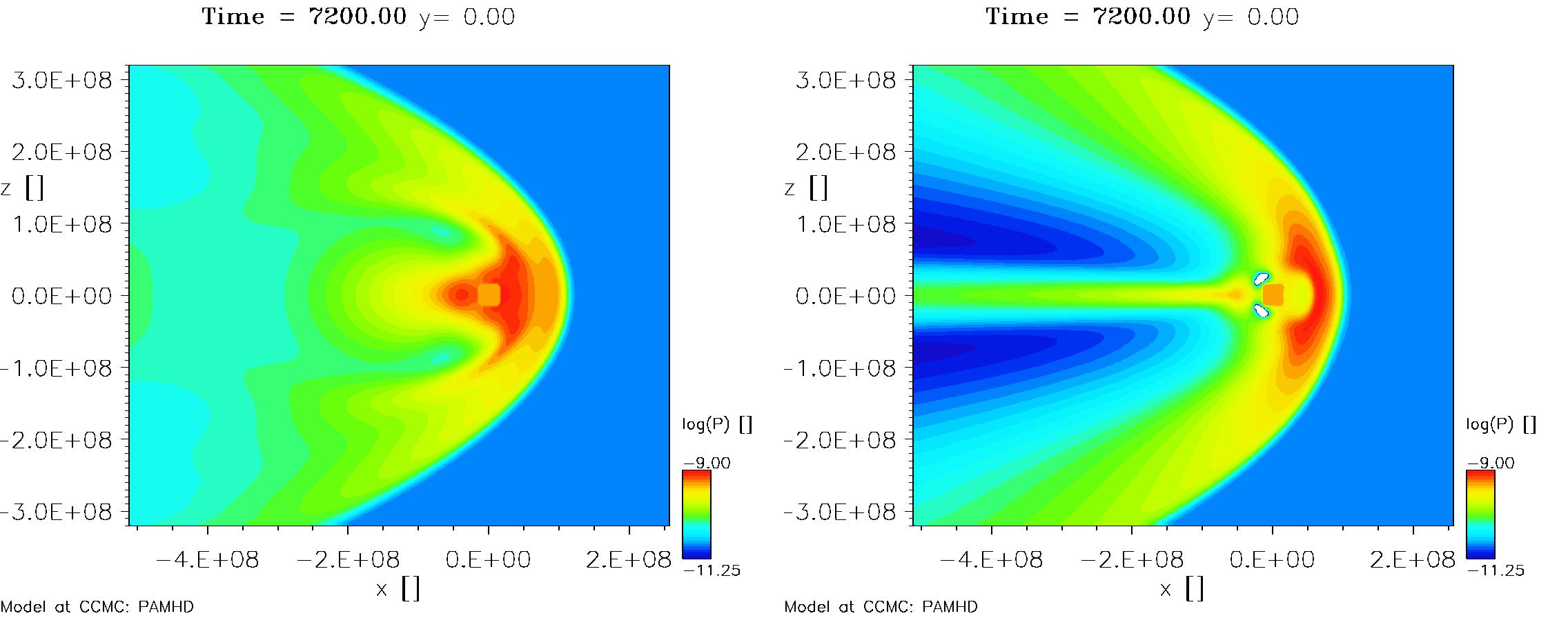}
\caption{
Pressure in simulation of interaction of Earth's magnetosphere with solar wind.
Left: northward (up) interplanetary magnetic field (IMF) plotted from http://ccmc.gsfc.nasa.gov/results/viewrun.php?domain=LP\&runnumber=ilja\_honkonen\_20161017\_LP\_4b.
Right: southward (down) IMF plotted from http://ccmc.gsfc.nasa.gov/results/viewrun.php?domain=LP\&runnumber=ilja\_honkonen\_20161020\_LP\_1a
}
\label{fig:earth}
\end{figure}

Figure \ref{fig:earth} shows pressure in three-dimensional simulations of Earth's magnetosphere with setup similar to \cite{powell99}:
incoming solar wind flow is modeled as a value boundary with speed of 400 km/s parallel to X axis, density is 5 protons/cm$^3$, thermal pressure 12.54 pPa and magnetic field 10 nT parallel (southward, on the right) or anti-parallel (northward, on the left) to Earth's dipole moment of $-7.94 \times 10^{22} A m^2$ along Z axis.
Inner boundary of the simulation is created at 3 R$_E$ with value boundaries of zero velocity and magnetic field, density 1000 protons/cm$^3$ and a copy boundary for pressure.
We use the Rusanov solver and 0.5 $R_E$ resolution where $R_E = 6371$ km.
Run ids of these examples are ilja\_honkonen\_20161020\_LP\_1a and 20161017\_LP\_4b.
Both the bowshock and magnetopause are located 1.5 $R_E$, or 3 grid cells, closer to Earth when IMF is southward, i.e. anti-parallel to planetary field on dayside magnetopause, versus northward.

As another example of the flexibility offered by PAMHD, we present to our knowledge the first MHD simulation of the interaction of magnetospheres of Jupiter and Saturn in two dimensions that was entirely prepared, executed and plotted through the CCMC RoR website.
The possibility of Saturn and Voyager 2 traversing Jupiter's wake was reported in \cite{scarf79} but at the time investigating this interaction using MHD was not feasible.
PAMHD can be quickly used to look at this unusual but real case as the configuration is otherwise the same as for Earth's magnetosphere but with two planets within the simulation box instead of one.

\begin{figure}
\includegraphics[width = \columnwidth]{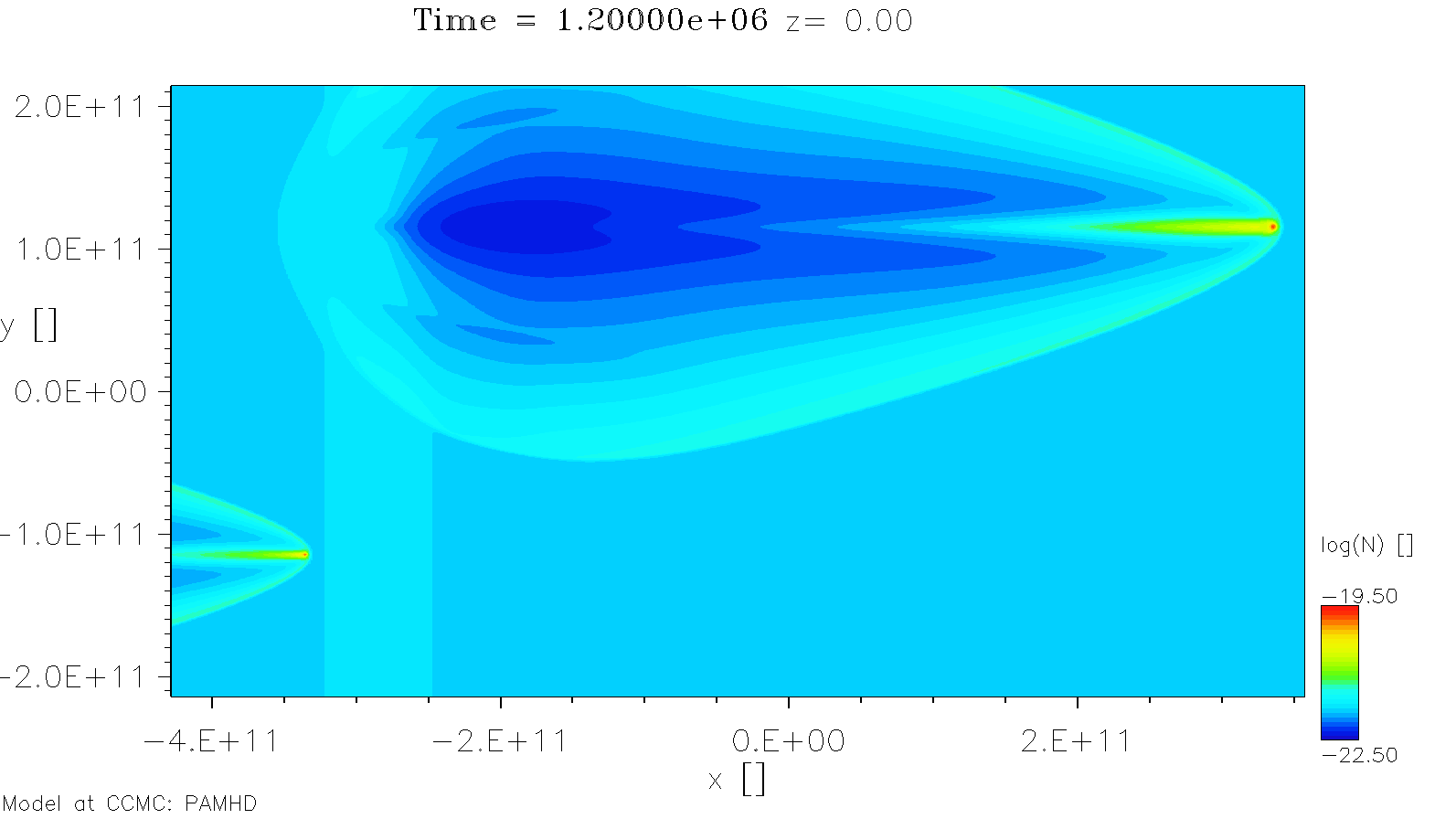}
\caption{
Mass density at approx 14 days from start of 2-dimensional simulation of interaction of magnetospheres of Jupiter and Saturn plotted from http://ccmc.gsfc.nasa.gov/results/viewrun.php?domain=LP\&runnumber=ilja\_honkonen\_20161021\_LP\_3
}
\label{fig:jupsat}
\end{figure}

Figure \ref{fig:jupsat} shows mass density in a 2-dimensional simulation of the interaction of magnetospheres of Jupiter and Saturn at approximately 14 days from start of simulation.
We use Rusanov solver, 4400 by 2400 cells excluding boundaries giving a resolution of 2.5 $R_J$ where $R_J = 71372$ km.
The basic setup is based on orbits and solar wind parameters reported in \cite{lepping83}:
Saturn orbit is 9400 $R_J$ further from the Sun than Jupiter's, the simulation is initialized with solar wind flowing along horizontal axis with speed 400 km/s, proton density is 0.1/cm$^3$ with 20\% sinusoidal variation starting at $2 \times 10^6 s \approx 23 d$ with period of 25 days, magnetic field magnitude 0.5 nT anti-parallel to planetary dipole directions and thermal pressure 21 pPa.
We create the inner boundary representing both planets at a distance of 10 $R_J$ from their center and similarly to Earth we use value boundaries for density, velocity and magnetic field, and copy boundaries for pressure.
We use dipole strengths of $1.43 \times 10^{27} A m^2$ for Jupiter and $4.61 \times 10^{25} A m^2$ for Saturn based on scalings relative to Earth given in \cite{olson06}.
Run id of this example is ilja\_honkonen\_20161021\_LP\_3.

\begin{figure}
\includegraphics[width = \columnwidth]{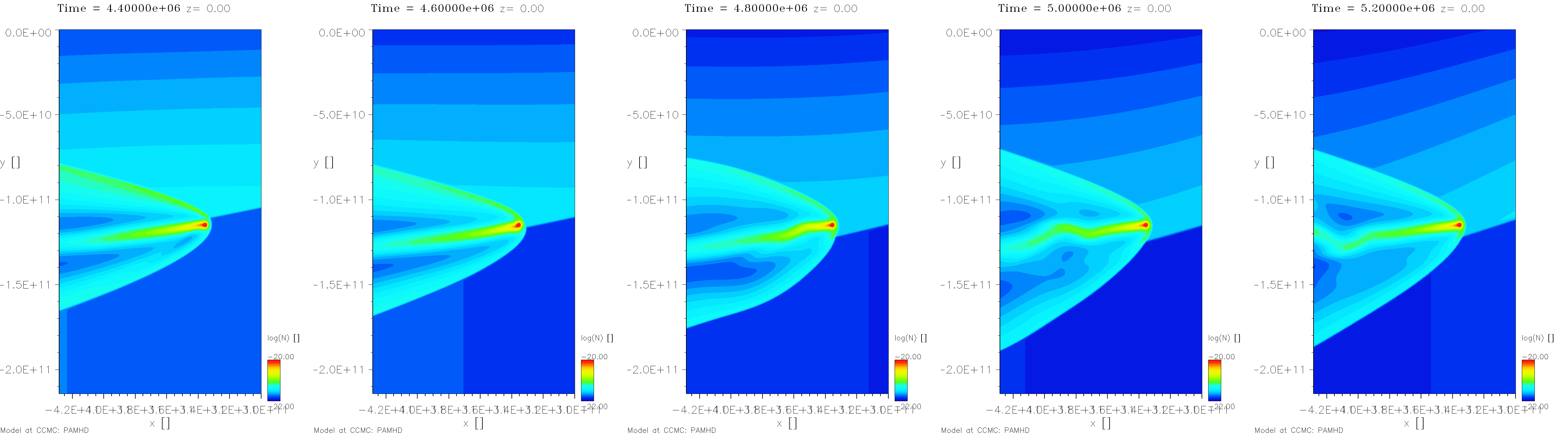}
\caption{
As Figure \ref{fig:jupsat} but showing a time series view of Saturn at approximately 2.3 day intervals.
The extent of each snapshot is approximately $2800 R_J$ in vertical and $1680 R_J$ in horizontal direction.
}
\label{fig:jupsat2}
\end{figure}

Figure \ref{fig:jupsat2} shows, at approximately 2.3 day intervals, the effect of solar wind density variation on the motion of Jupiter's bowshock and magnetosheath and the resulting effect on Saturn's magnetosphere.
The passage of Jupiter's magnetosheath over Saturn leads to a significant perturbation of Saturn's magnetotail.
Saturn's neutral sheet visible in Figure \ref{fig:jupsat2} moves up to  $90 R_J$ perpendicular to Sun-Saturn line just as Jupiter's bowshock crosses over Saturn's bowshock.

\section{Future developments}

As the presented model is free and open source software, but still lacks basic functionality for some areas of helio and astrophysics, it is well suited for collaboration and contributions from many levels of expertise.
The following is an incomplete list of topics and references which could be implemented and studied by pre-grad, post-grad or post-doc scientists.
The difficulty of these tasks can vary widely so we are happy to discuss the details with interested parties.
\begin{itemize}
\item Reflecting boundary condition in which scalars are copied and vectors are mirrored with respect to boundary geometry
\item Adaptive mesh refinement, already supported by the grid library used by PAMHD \cite{honkonen13}.
\item Limiters for 2nd order and higher quality solutions \cite{mccorquodale11}.
\item Constrained transport of magnetic field \cite{londrillo04} to preserve $\nabla\cdot B = 0$ to numerical accuracy during MHD solution
\item A source term for plasma representing charge exchange with neutral particles e.g. in outer parts of heliosphere \cite{schwadron11} which slows down and heats the solar wind.
\item Interactions of plasma and dust \cite{fahr85}.
\item Chemical reactions, relevant not only for dust-plasma interactions but also for studying ionospheric outflow from weakly or non-magnetized planets \cite{ma13}.
\item Self-gravity of plasma, relevant for e.g. star and galaxy formation, using the existing Poisson equation solver used for cleaning $\nabla\cdot B$.
\item Ionospheric boundary condition, using e.g. an empirical model for electric potential \cite{weimer95}, for a more self-consistent global MHD simulation.
\end{itemize}

\section{Conclusions}
\label{sec:conclusions}

We present a new magnetohydrodynamic (MHD) model that anyone can download, use, study, modify and redistribute.
The model is implemented using C++11 and supports an arbitrary number of initial and boundary conditions for simulation variables in box and sphere geometries.
Furthermore the model is available through NASA's Community Coordinated Modeling Center (CCMC) Runs on Request (RoR) system which enables anyone to request simulations and visualize the results through CCMC's website.
Using RoR we present several simulations of shocks and instabilities as well as the first to our knowledge MHD simulation of the interaction of magnetospheres of Jupiter and Saturn.
A periodic 20\% variation in solar wind density causes significant changes in Saturn's magnetosphere as it periodically enters and exits Jupiter's magnetosheath.
A possibly even larger effect could be caused by a small (e.g. 10 degree) periodic variation in solar wind direction that would

\section{Acknowledgments}
\label{sec:acks}

Most of this work was performed while IH was funded through the NASA Postdoctoral Program administered by Oak Ridge Associated Universities.
We a grateful to developers/contributors of following programs/libraries for making them freely available: athena \cite{stone08}, boost, cxx-prettyprint, eigen, muparserx, rapidjson and zoltan \cite{devine02}.

\bibliographystyle{elsarticle-num}
\bibliography{references}

\end{document}